\documentclass[sigconf,nonacm]{acmart}
\renewcommand\footnotetextcopyrightpermission[1]{} 
\AtBeginDocument{%
  }

\usepackage{amsmath}
\usepackage{amsthm}
\usepackage{amssymb}
\usepackage{hyperref}
\usepackage[capitalise]{cleveref}
\usepackage{xfrac} 
\usepackage{xspace} 
\usepackage{booktabs}
\usepackage{multirow}
\usepackage[ruled,vlined,linesnumbered]{algorithm2e}
\usepackage{caption}
\usepackage{subcaption}
\usepackage{chato-notes}
\usepackage{MnSymbol}
\usepackage[official]{eurosym}
\usepackage{siunitx}
\usepackage{float}
\usepackage{url}

\newtheorem{lemma}{Lemma}

\newcommand{\spara}[1]{\smallskip\noindent{\bf #1}}

\definecolor{verylightgray}{rgb}{0.8,0.8,0.8}
\definecolor{brown}{rgb}{0.6, 0.29, 0.0}
\definecolor{forestgreen}{rgb}{0.0, 0.5, 0.0}
\definecolor{blue-violet}{rgb}{0.54, 0.17, 0.89}
\definecolor{dartmouthgreen}{rgb}{0.05, 0.5, 0.0}
\definecolor{dark-red}{rgb}{0.7, 0.0, 0.0}

\usepackage{xcolor}

\usepackage{enumitem} 

\RestyleAlgo{ruled}
\SetKwInput{KwInput}{Input}
\SetKwInput{KwOutput}{Output}

\usepackage{booktabs}
\usepackage{makecell}

\usepackage{cprotect} 
\newcommand{\our}{\texttt{WRSWR-SKIP} }

\begin{document}

\title{Weighted Reservoir Sampling With Replacement from Data Streams}


\author{Adriano Meligrana}
\authornote{Corresponding author.} 
\affiliation{%
  \institution{Sapienza University of Rome}
  \city{Rome}
  \country{Italy}}
\email{adriano.meligrana@diag.uniroma1.it}

\author{Adriano Fazzone}
\affiliation{%
  \institution{Intesa Sanpaolo Innovation Center}
  \city{Turin}
  \country{Italy}}
\email{adriano.fazzone@intesasanpaolo.com}

\renewcommand{\shortauthors}{Meligrana and Fazzone}

\begin{abstract}
In this work, we present a new random sampling method for data streams where the probability of an element's inclusion in the sample is proportional to a weight associated with that element. 
Our method is based on sampling with replacement, although most of the literature on this topic has focused on sampling without replacement.

Our algorithm generates a weighted random sample in one pass over a population of unknown size. At any point in time, the sample is representative of the population seen so far and can be directly used by other modules without requiring any post-processing.
We formally prove the correctness and efficiency of our method. 

An experimental analysis shows the performance of our method in practice when compared to state-of-the-art methods.
\end{abstract}

\keywords{Reservoir Sampling, Random Sampling, Weighted Sampling, Data Streams}

\maketitle

\section{Introduction}
\label{sec:introduction}

One fundamental problem in several computing fields, including databases and data mining, is the efficient summarization of massive datasets and data streams \cite{muthukrishnan2005data, cormode2020small}. Random sampling serves as an essential tool in this environment, effectively reducing large volumes of data to a manageable subset while preserving essential statistical properties \cite{chaudhuri1999random, shekelyan2023EDBT}. This problem is particularly important in the streaming setting, where data arrive sequentially at a high rate and the total population size is often unknown.

The classical approach for sampling from unknown-sized populations is reservoir sampling  \cite{park2004SDM, park2007computational, vitter1985random}. Extensions of this model to weighted streams have largely concentrated on selection without replacement, where selected elements must be distinct \cite{efraimidis2006information}. This has led to significant work on efficient, one-pass algorithms for weighted reservoir sampling without replacement (WRSWOR) \cite{efraimidis2006information, braverman2019weighted, rajesh2019weighted}.

However, there are important applications where reservoir sampling with replacement (RSWR) is necessary or strongly preferred, particularly for statistical estimation tasks that rely on the independence of sampled elements \cite{park2007computational}, a property that samples without replacement do not possess. While a sample without replacement can be transformed to satisfy this independence requirement, doing so incurs additional computational costs that direct RSWR avoids. Weighted reservoir sampling with replacement (WRSWR) is essential when uniform selection is inadequate, enabling sophisticated algorithms such as the weighted bootstrap and approximate query processing in the streaming setting \cite{park2007computational, chaudhuri1999random, shekelyan2023EDBT, kojadinovic2012goodness}. While conceptual reductions from weighted to unweighted RSWR have been noted (e.g., duplicating an element $w$ times \cite{rajesh2019weighted}), the literature remains heavily focused on the without-replacement case~\cite{efraimidis2006information, braverman2019weighted, rajesh2019weighted}. 
Our work addresses this relative scarcity of specialized methods by presenting a novel and efficient approach focused on sampling with replacement.

This paper introduces a WRSWR method designed specifically for data streams. Our algorithm generates a weighted random sample in one pass over a population of unknown size. A critical feature of this algorithm is that at any point in time, the sample is representative of the population seen thus far, allowing it to be used immediately by other modules without requiring any post-processing.

\section{Related Work}
\label{sec:related_work}
Weighted reservoir sampling with replacement was addressed in the pioneering work of Chaudhuri et al.~\cite{chaudhuri1999random}. The authors presented an algorithm that uses Bernoulli trials to continuously maintain a weighted reservoir sample of the stream; we refer to this algorithm as \texttt{WRSWR} throughout this paper.

We also define \texttt{WRSWR-BIN} as the improvement to \texttt{WRSWR}, first described in Park et al.~\cite{park2007computational}, which substitutes the series of Bernoulli trials with a single Binomial experiment. As we show in Section~\ref{sec:experiments}, this optimization makes the method practical to use.

However, these algorithms do not implement any weight skipping techniques~\cite{efraimidis2006information, vitter1985random, park2007computational}, making their performance suboptimal. A key contribution of our work is the correct adaptation of this technique for the weighted case.

Shekelyan et al.~\cite{shekelyan2023EDBT} presented an online multinomial sampler algorithm based on an adaptation of the sampling technique from Efraimidis and Spirakis~\cite{efraimidis2006information}. We refer to this algorithm as \texttt{WRAExp-J} throughout this paper.

However, this method requires transforming a weighted sample without replacement into a weighted sample with replacement when retrieving the reservoir, a weakness that our algorithm avoids.

To the best of our knowledge, there are no other methods that directly address the weighted reservoir sampling with replacement problem.

\section{Our Method}
\label{sec:our_method}
We first define the notation used in this paper. 
Let $\mathcal{S}$ be a stream of items, where each item $\mathcal{S}_t = (e_t, w_t)$ for $t \in \mathbb{Z}_{\geq 1}$ is a pair consisting of a stream element $e_t$ and its corresponding weight $w_t \in \mathbb{R}_{>0}$. 
We denote the reservoir as $\mathcal{R}$, a data structure of fixed size $m$ that stores stream elements in positions indexed from 1 to $m$.

Our contribution is \our, a weighted reservoir sampling with replacement algorithm, which we present in Algorithm~\ref{alg:our}.
\begin{algorithm}[h!]
	\caption{\our}
	\label{alg:our}
	\KwInput{$\mathcal{S}$, $N$, $m$.}
	$(e_1,w_1) \gets \mathcal{S}_1$\;
	$\mathcal{R} \gets [\underbrace{e_1, e_1, \dots, e_1}_{m \text{ times}}]$;
    
    $W \gets w_1$\;
    $q \sim U(0,1)$\;
    $W_{\text{skip}} \gets W /q ^{1/m}$\;
	\For{$t = 2$ \KwTo $N$}{
		$(e_t, w_t) \gets \mathcal{S}_t$\;
		$W \gets W + w_t$\;
		\If{$W \geq W_{\text{skip}}$}{
			$q \sim U(0,1)$\;
			$W_{\text{skip}} \gets W/q^{1/m}$\;
			$k \sim B_{>0}\left(m,\frac{w_t}{W}\right)$\;
			Insert $e_t$ in $k$ distinct positions in $\mathcal{R}$ chosen uniformly at random\;
		}
	}
	\Return $\mathcal{R}$\;
\end{algorithm}

Algorithm~\ref{alg:our} begins by initializing the reservoir $\mathcal{R}$ to contain $m$ copies of the first stream element $e_1$ (Line 2) and setting the initial cumulative weight $W$ to $w_1$ (Line 3). The initial skip threshold, $W_{skip}$, is then calculated by drawing a uniform random number $q \sim U(0,1)$ (Line 4) and setting $W_{skip} \leftarrow W/q^{1/m}$ (Line 5).

The algorithm then iterates through the stream from $t=2$ to $N$ (Line 6). In each iteration, the current item $(e_t, w_t)$ is retrieved (Line 7), and its weight $w_t$ is added to the cumulative total weight $W$ (Line 8).

The core of the algorithm's efficiency lies in its skipping mechanism. The algorithm only considers an update if the total weight $W$ has reached or surpassed the $W_{skip}$ threshold (Line 9). This check efficiently simulates the cumulative probability of rejecting a sequence of items without processing each one individually.

If this condition is met, the algorithm proceeds to update the reservoir. First, it determines the next skip threshold by drawing a new uniform random number $q \sim U(0,1)$ (Line 10) and setting $W_{skip} \leftarrow W/q^{1/m}$ (Line 11). This new threshold defines the minimum total weight that must be accumulated before the next item can be processed.

Next, the number of copies $k$ for $e_t$ to be inserted into the reservoir is drawn from $B_{>0}\left(m, \frac{w_t}{W}\right)$, a Binomial distribution truncated at zero (Line 12). The truncation guarantees that $e_t$ replaces at least one element ($k>0$) in $\mathcal{R}$.

Finally, the element $e_t$ is inserted into $k$ distinct positions in $\mathcal{R}$, chosen uniformly at random (Line 13). The algorithm then continues to the next item in the stream. If $N$ items have been read, it returns the final reservoir $\mathcal{R}$ (Line 14).

We now prove the correctness of the \our algorithm.

\begin{lemma}\label{lemma:our_correctness}
    Algorithm~\ref{alg:our} keeps an unbiased weighted random sample with replacement at each iteration.
	\begin{proof}
Consider an unoptimized version of Algorithm~\ref{alg:our} that for $t=1$ fills the reservoir $\mathcal{R}$ of size $m$ with $e_1$ and for each $2 \leq t \leq N$ draws a number $k_t \sim B(m, w_t / W)$ and replaces $k_t$ elements uniformly at random from $\mathcal{R}$. 

We will prove by induction on $N$ that $P(\mathcal{R}[j] = e_i) = w_i / W_N$ for all $i \le N$. 
If $N=1$, $\mathcal{R}$ is filled with $m$ copies of $e_1$ and the total weight is $W_1 = w_1$, and $P(\mathcal{R}[j] = e_1) = w_1/W_1 = 1$; then the base case holds. 

Assume that after processing $N$ items, the property holds: 
for any slot $j\in [m]$ and any $i \le N$ we have that $P(\mathcal{R}[j] = e_i \text{ at step } N) = \frac{w_i}{W_N}$. 

We now process item $(e_{N+1}, w_{N+1})$.  The total weight is updated to $W_{N+1} = W_N + w_{N+1}$ and the replacement probability is $p = w_{N+1} / W_{N+1}$. 
Let $K \sim B(m, p)$ be the random variable for the total number of slots to be replaced. We are interested in the probability that a specific slot $j$ is replaced by $e_{N+1}$.Given a specific outcome $K=k$, these $k$ slots are chosen uniformly at random from the $m$ available slots. Therefore, the conditional probability that slot $j$ is included in this selection, given $K=k$, is $k/m$.
Therefore, the conditional probability of replacement for slot $j$, given $k$, is $P(\mathcal{R}[j] \gets e_{N+1} | K=k) = k/m$. By the law of total probability $P(\mathcal{R}[j] \gets e_{N+1}) = \sum_{k=1}^m P(K=k)[P(\mathcal{R}[j] \gets e_{N+1} | k)] = \frac{1}{m} \sum_{k=1}^m P(K=k)\cdot k = \mathbb{E}[K]/m = p = \frac{w_{N+1}}{W_{N+1}}$. 
Therefore for the new item $e_{N+1}$, $P(\mathcal{R}[j] = e_{N+1} \text{ at } N+1) = P(\mathcal{R}[j] \gets e_{N+1}) = \frac{w_{N+1}}{W_{N+1}}$. 
An old item $e_i$ where $i \leq N$, is in slot $j$ only if it was there at step $N$ and the slot was not replaced at step $N+1$.
\small
\[
P(\mathcal{R}[j] = e_i \text{ at } N+1) =P(\mathcal{R}[j] = e_i \text{ at } N) ~ P(\mathcal{R}[j] \text{ is not replaced at } N+1).
\]
\normalsize
Using the induction hypothesis and our probability from above we have that:
\small
\[
P(\mathcal{R}[j] = e_i \text{ at } N+1) = \left( \frac{w_i}{W_N} \right)  \left( 1 - \frac{w_{N+1}}{W_{N+1}} \right) = \left( \frac{w_i}{W_N} \right) \left(\frac{W_{N}}{W_{N+1}} \right) = \frac{w_i}{W_{N+1}}.
\]
\normalsize
Since the property holds for $N+1$, the unoptimized algorithm is correct by induction.

Now, let's focus on the fact that an update of $\mathcal{R}$ is only needed if $k_t >0$. 
Let $\tau$ be the step of the last update of $\mathcal{R}$. The algorithm will skip all steps from $\tau+1$ to $t-1$ and then it first performs an update of $\mathcal{R}$ at step $t$ with probability:
\small
\begin{align}
&P(\text{first update of $\mathcal{R}$ at } t) = P(k_{\tau+1}=0, \dots, k_{t-1}=0, k_t > 0) \nonumber\\
&= \left[ \prod_{j=\tau+1}^{t-1} P(k_j=0) \right] P(k_t > 0) 
= \left[ \prod_{j=\tau+1}^{t-1} \left(\frac{W_{j-1}}{W_j}\right)^m \right] \left( 1 - \left(\frac{W_{t-1}}{W_t}\right)^m \right) \nonumber\\
&= \left(\frac{W_\tau}{W_{t-1}}\right)^m \left( 1 - \left(\frac{W_{t-1}}{W_t}\right)^m \right)
= \left(\frac{W_\tau}{W_{t-1}}\right)^m - \left(\frac{W_\tau}{W_t}\right)^m.
\end{align}
\normalsize
At step $\tau$, Algorithm~\ref{alg:our} sets $W_{\text{skip}} = W_\tau / q^{1/m}$ where $q \sim U(0,1)$. The next update will be at the \emph{first} step $t > \tau$ where $W_t \ge W_{\text{skip}}$.
The probability that this first update step is at $t$ is:
\small
\begin{align*}
&P(\text{first update of $\mathcal{R}$ at } t) = P(W_{t-1} < W_{\text{skip}} \le W_t) \\
&= P\left(W_{t-1} < \frac{W_\tau}{q^{1/m}} \le W_t\right) 
= P\left( \frac{1}{W_{t-1}} > \frac{q^{1/m}}{W_\tau} \ge \frac{1}{W_t} \right) \\
&= P\left( \frac{W_\tau}{W_{t-1}} > q^{1/m} \ge \frac{W_\tau}{W_t} \right)
= P\left( \left(\frac{W_\tau}{W_{t-1}}\right)^m > q \ge \left(\frac{W_\tau}{W_t}\right)^m \right).
\end{align*}
\normalsize
Since $q$ is uniform on $(0,1)$, the probability of $q$ falling into this interval is simply the length of the interval:
\begin{align}
P(\text{first update of $\mathcal{R}$ at } t) = \left(\frac{W_\tau}{W_{t-1}}\right)^m - \left(\frac{W_\tau}{W_t}\right)^m.
\end{align}
Since the probabilities in (1) and (2) are identical, Algorithm~\ref{alg:our} is correct.

	\end{proof}
\end{lemma}

Following \cite{efraimidis2006information, park2007computational,vitter1985random}, we evaluate the efficiency of a reservoir sampling algorithm based on the required number of random variates.
\begin{lemma}\label{lemma:our_complexity}
    Algorithm~\ref{alg:our} requires $O(m \log \frac{W_N}{w_1})$ random variates in expectation to process the first $N$ elements of a stream.
	\begin{proof}
        Let $X$ be the random variable for the total number of random variates generated while processing the first $N$ elements. 
        We can separate this total into the initialization phase (lines 1-5) and the loop phase (lines 6-13). 
        In the initialization phase, only one random variate is drawn at line 4. For the loop phase, let $X_t$ be the number of random variates generated during the $t$-th iteration (where $t \in [2, N]$). Random variates are only generated if the condition on line 9 is met. Let $I_t$ be the indicator variable that the conditional block is entered at step $t$. $P(I_t=1) = 1 - \left(\frac{W_{t-1}}{W_{t}}\right)^m$ since this is the probability of accepting $e_t$ at least once. If the block is not entered ($I_t=0$), the number of variates generated is $0$. If the block is entered ($I_t=1$), the algorithm requires a random variate for the draw $q \sim  U(0,1)$, another random variate to draw $k$ from $K_t \sim B_{>0}(m, \frac{w_t}{W_t})$, and $k$ variates to replace the elements in the reservoir. The total expectation of number of random variates $\mathbb{E}[X]$ is then equal to $1 + \sum_{t=2}^N \mathbb{E}[X_t]$. By the previous considerations, $\mathbb{E}[X_t] = \mathbb{E}[I_t](2+\mathbb{E}[K_t])$, therefore 
    \small
    \[
    \mathbb{E}[X] = 1+\sum_{t=2}^N \mathbb{E}[I_t](2+\mathbb{E}[K_t]) = 1+\sum_{t=2}^N \left(1-\left(\frac{W_{t-1}}{W_{t}}\right)^m\right)\left(2 + m\frac{W_{t} - W_{t-1}}{W_{t}} \right).
    \]
    \normalsize
    	Notice that 
        \small
    	\begin{align*} 
    		\mathbb{E}[X] -1 \leq  3 m \sum_{t=2}^N \left(\frac{W_{t} - W_{t-1}}{W_{t}}\right) < 3m\sum_{t=2}^N \ln \frac{W_{t}}{W_{t-1}} = 3m\ln \frac{W_{N}}{w_1}.
    	\end{align*} 
        \normalsize
    	The first inequality follows from the fact that, if we define \(x := \tfrac{W_{t-1}}{W_{t}} \in \mathbb{R}_{(0,1)}\), the factor \((1 - x^m)(2 + m(1 - x))\) is maximized by \(3m(1 - x)\), while the second inequality follows by the convexity of the $\ln$ function. 
        Therefore, $\mathbb{E}[X]= O\left(m \log \frac{W_N}{w_1}\right)$ as stated.
	\end{proof}
\end{lemma}


For clarity of exposition, we define two primary operations: \texttt{Add} and \texttt{Get}. The \texttt{Add} operation refers to processing a single item from the stream, which is analogous to a reservoir update. The \texttt{Get} operation refers to obtaining a sample of the entire input stream from the algorithm.


We evaluate the performance of \our, in terms of its \texttt{Add} and \texttt{Get} operations, against the following three algorithms presented in Section~\ref{sec:related_work}: \texttt{WRSWR}~\cite{chaudhuri1999random}, \texttt{WRSWR-BIN}, and \texttt{WRAExp-J}~\cite{shekelyan2023EDBT}.

The \texttt{Add} complexity for \texttt{WRAExp-J} is presented in Efraimidis et al.~\cite{efraimidis2006information} and holds under the i.i.d. assumption for the weight distribution in the stream.
The $O(m)$ complexity for the \texttt{Get} operation is evident from inspecting the loop at lines 4-9 in Algorithm 2 of Shekelyan et al.~\cite{shekelyan2023EDBT}.

The \texttt{Get} complexity is $O(1)$ for both \texttt{WRSWR} and \texttt{WRSWR-BIN} because their reservoirs do not require any post-processing to generate a weighted random sample from the input stream.

The $O(Nm)$ \texttt{Add} complexity for \texttt{WRSWR} arises from a reduction of the algorithm by Chao~\cite{chao1982general}, which is equivalent to maintaining $m$ independent reservoirs of size 1.
 
\texttt{WRSWR-BIN} coincides with the unoptimized algorithm presented at the beginning of the proof of Lemma~\ref{lemma:our_correctness}, and the cost of its \texttt{Add} operation has an additional $O(N)$ term compared to the cost of the same operation for the \texttt{WRSWR-SKIP} algorithm.

\begin{table}[htbp]
\centering
\small 
\begin{tabular*}{\columnwidth}{@{\extracolsep{\fill}} l cccc} 
\toprule
\textbf{Op.} & \textbf{\texttt{WRSWR}} & \textbf{\texttt{WRSWR-BIN}} & \textbf{\texttt{WRAExp-J}} & \textbf{\texttt{WRSWR-SKIP}} \\
\midrule
\texttt{Add} & $O\left(Nm\right)$ & $O\left(N+m\log\frac{W_N}{w_1}\right)$ & $O\left(m\log \frac{N}{m}\right)^*$ & $O\left(m\log\frac{W_N}{w_1}\right)$ \\
\texttt{Get} & $O\left(1\right)$ & $O\left(1\right)$ & $O\left(m\right)$ & $O\left(1\right)$ \\
\bottomrule
\end{tabular*}
\caption{Time complexity (expected number of random variates) for the \texttt{Add} and \texttt{Get} operations. $^*$The \texttt{Add} complexity for \texttt{WRAExp-J}~\cite{efraimidis2006information} holds under the i.i.d. assumption for the weight distribution.}
\label{tab:time_complexity_comparison}
\vspace{-0.69cm}
\end{table}
Table~\ref{tab:time_complexity_comparison} highlights the main theoretical contribution of \texttt{WRSWR-SKIP}: it is the only algorithm listed that achieves both an optimal $O(1)$ \texttt{Get} complexity and an \texttt{Add} complexity that avoids linear dependence on the stream's length $N$, while remaining competitive with \texttt{WRAExp-J}.

\section{Experiments}
\label{sec:experiments}
We evaluate the performance of \our using both synthetic data and the Wikipedia Clickstream dataset\footnote{\url{https://meta.wikimedia.org/wiki/Research:Wikipedia_clickstream}}, benchmarking our proposed algorithm against the \texttt{WRSWR-BIN} and \texttt{WRAExp-J} baselines. 
We excluded \texttt{WRSWR} from our experiments due to its practical inefficiency. 
Moreover, we implemented the \texttt{Get} operation of \texttt{WRAExp-J} using a dynamic sampler \cite{zhang2023efficient, hafner2025exact} to improve its performance in practice. 
For both datasets, we report the average time over 100 runs for \texttt{add} and \texttt{get} operations, varying the reservoir size $m$ from $0.01\%$ to $10\%$ of the stream length $N$.
We implemented all algorithms in Julia 1.12.1, and the code is publicly available\footnote{\url{https://github.com/JuliaDynamics/StreamSampling.jl/tree/main/comparison}}. The experiments were run on a machine with an AMD Ryzen 5 5600H (6 cores) and 16GB of RAM, running Ubuntu 24.04 LTS.
\newcommand{\subfigwidth}{0.1561\textwidth}
\begin{figure}[h]
	\centering
	\begin{subfigure}[b]{\subfigwidth}
		\centering
		\includegraphics[width=\textwidth]{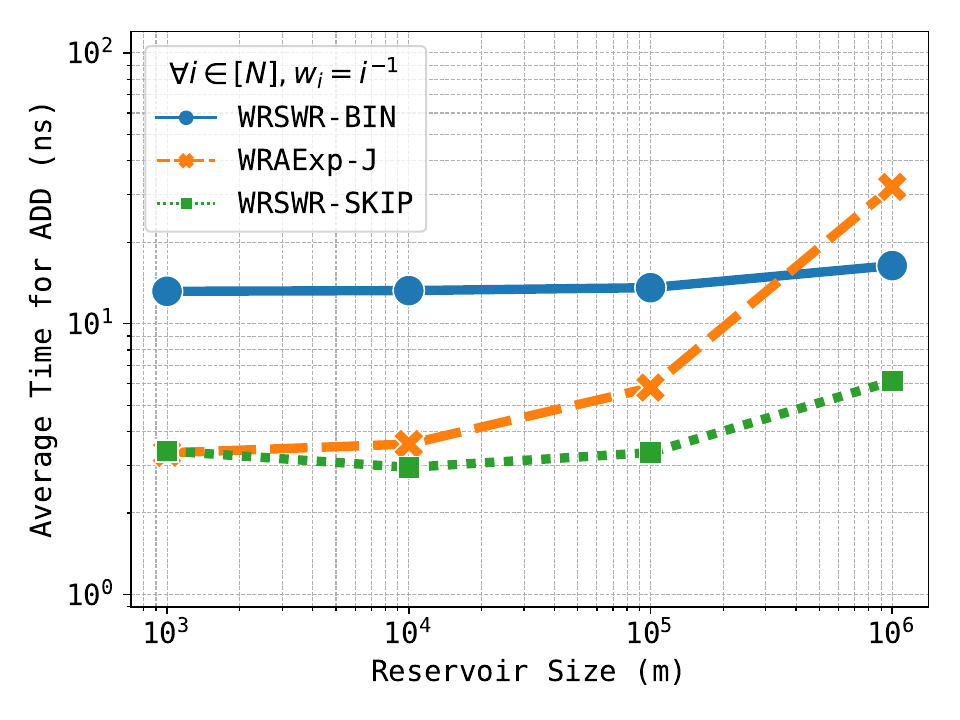}
		\caption{\tiny \texttt{Add}, decreasing weighs.}
		\label{fig:synthetic_experiment_add_dec}
	\end{subfigure}
	\hfill 
	\begin{subfigure}[b]{\subfigwidth}
		\centering
		\includegraphics[width=\textwidth]{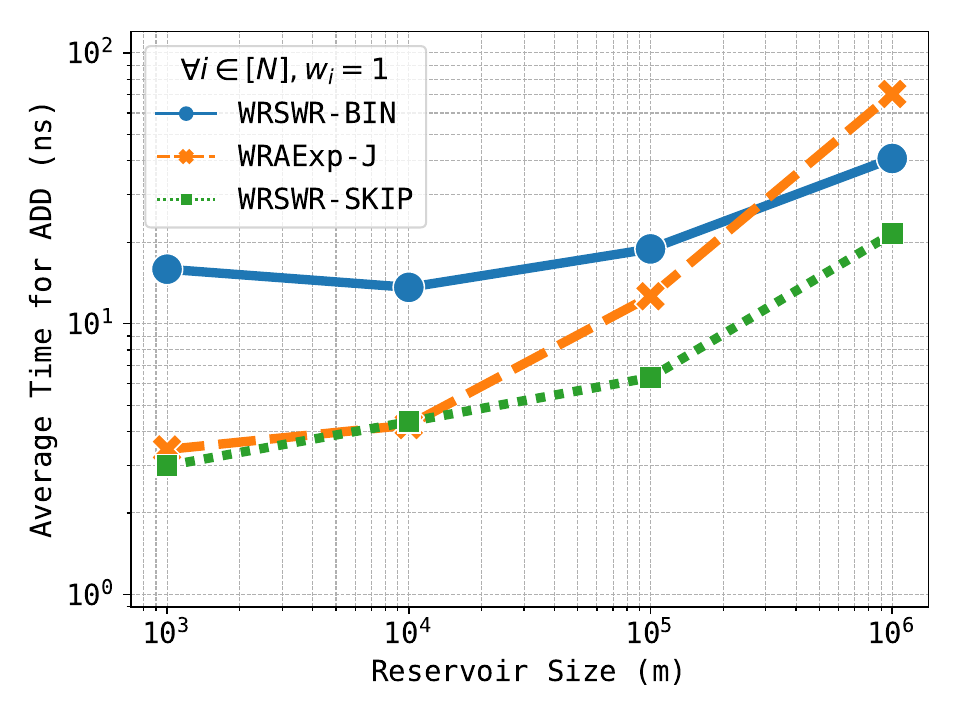}
		\caption{\tiny \texttt{Add}, constant weighs.}
		\label{fig:synthetic_experiment_add_const}
	\end{subfigure}
	\hfill 
	\begin{subfigure}[b]{\subfigwidth}
		\centering
		\includegraphics[width=\textwidth]{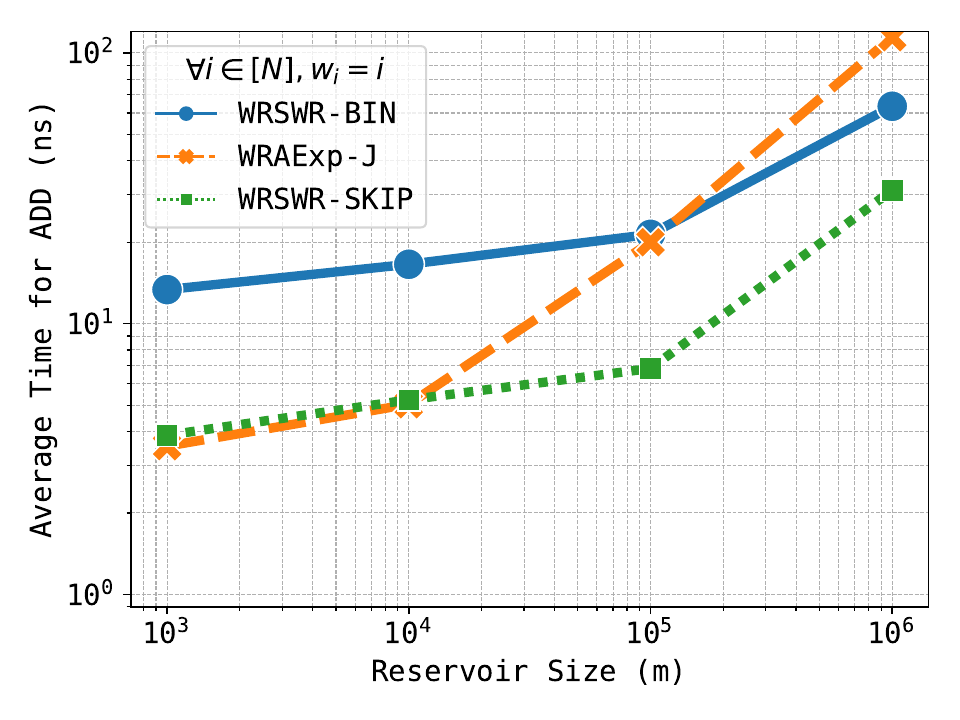}
		\caption{\tiny \texttt{Add}, increasing weighs.}
		\label{fig:synthetic_experiment_add_inc}
	\end{subfigure}
	    
	    
	\begin{subfigure}[b]{\subfigwidth}
		\centering
		\includegraphics[width=\textwidth]{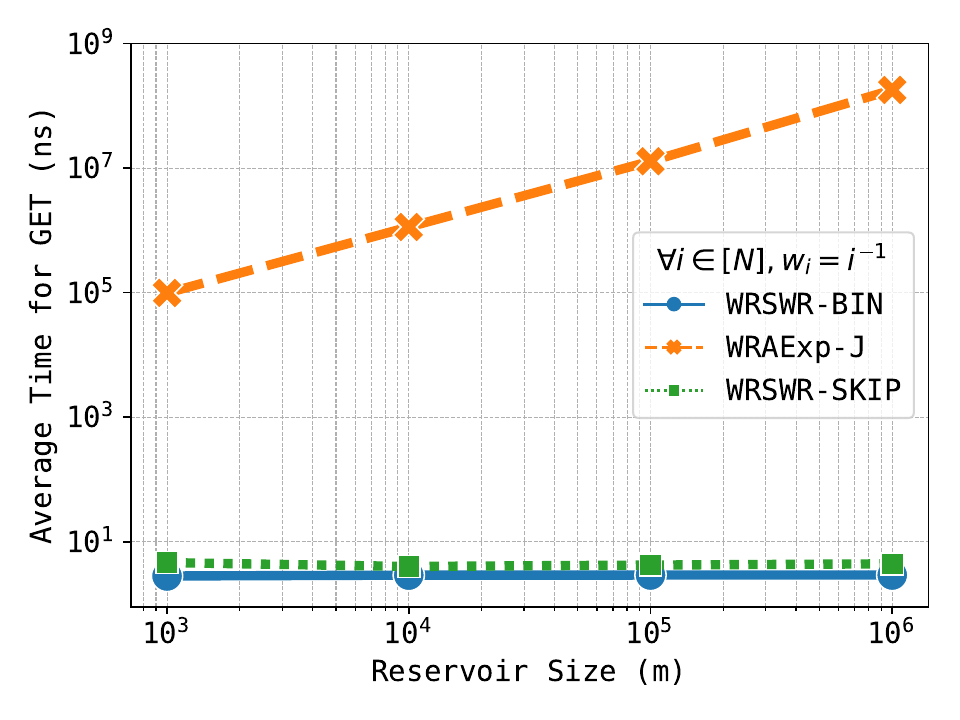}
		\caption{\tiny \texttt{Get}, decreasing weighs.}
		\label{fig:synthetic_experiment_get_dec}
	\end{subfigure}
	\hfill 
	\begin{subfigure}[b]{\subfigwidth}
		\centering
		\includegraphics[width=\textwidth]{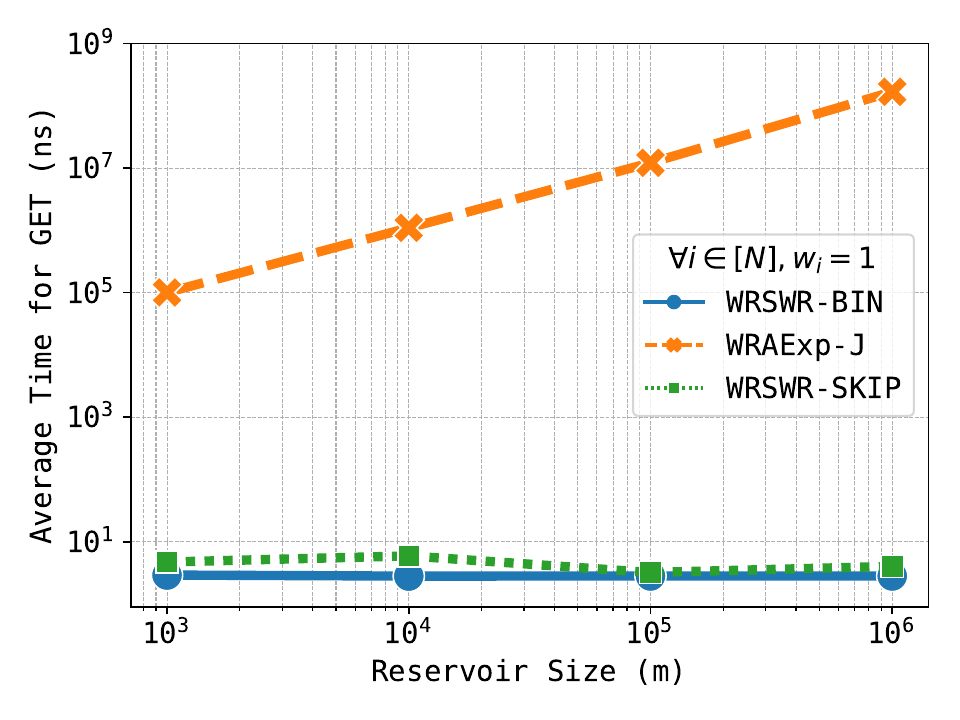}
		\caption{\tiny \texttt{Get}, constant weighs.}
		\label{fig:synthetic_experiment_get_const}
	\end{subfigure}
	\hfill 
	\begin{subfigure}[b]{\subfigwidth}
		\centering
		\includegraphics[width=\textwidth]{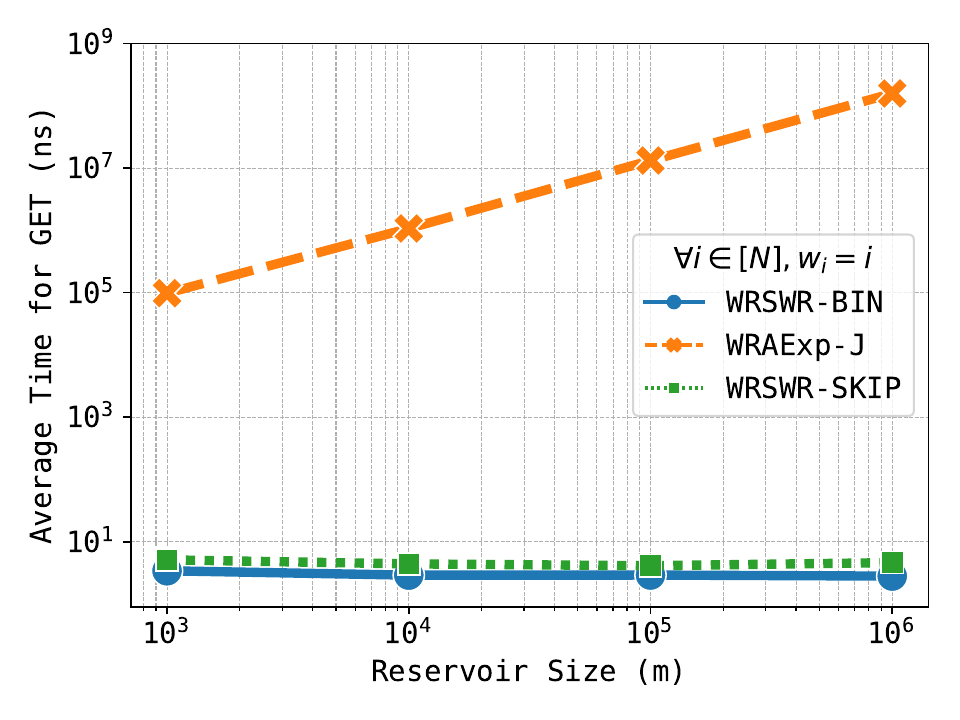}
		\caption{\tiny \texttt{Get}, increasing weighs.}
		\label{fig:synthetic_experiment_get_inc}
	\end{subfigure}
    
	\caption{\small Average time (ns) over 100 executions for \texttt{Add} (top) and \texttt{Get} (bottom) operations vs. reservoir size ($m$). Columns show performance on streams with decreasing, constant, and increasing weights, respectively. 10M items in the streams.}
	\label{fig:synthetic_experiment}
\end{figure}
\newcommand{\subfigwidthH}{0.236\textwidth}
	    
	
\begin{figure}[h]
    \centering
    \vspace{-0.3cm} 
    
    \begin{subfigure}[b]{\subfigwidthH}
        \centering
        \includegraphics[width=\textwidth, trim=0 0 0 0, clip]{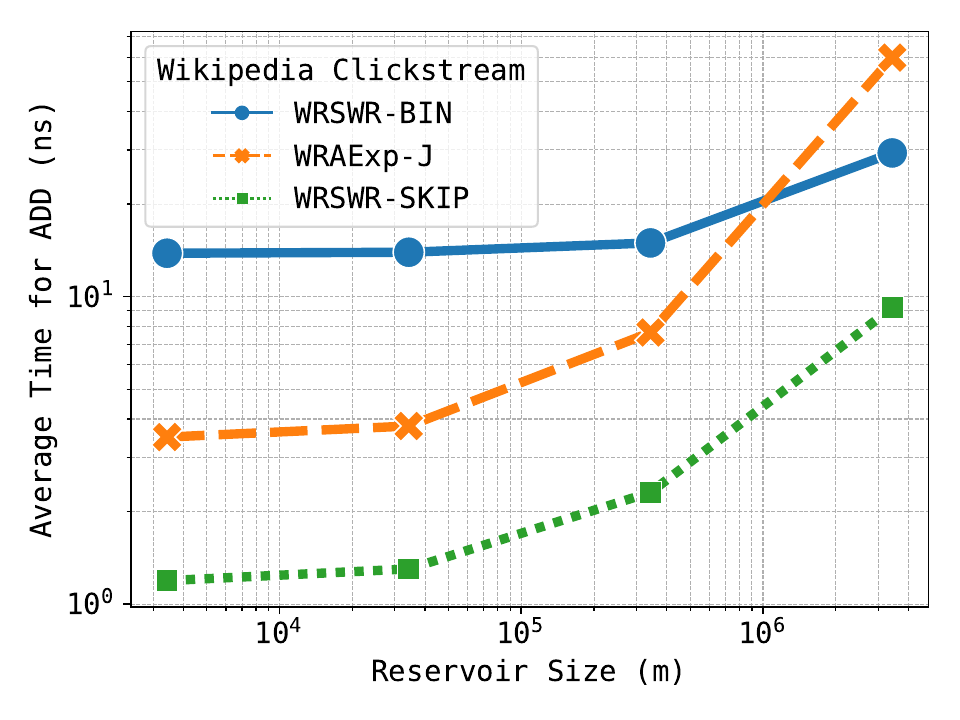}
        
        \vspace{-0.04cm}
        \caption{\small \texttt{Add}.}
        \label{fig:semisynthetic_experiment_add}
    \end{subfigure}
    \hfill 
    \begin{subfigure}[b]{\subfigwidthH}
        \centering
        \includegraphics[width=\textwidth, trim=0 0 0 0, clip]{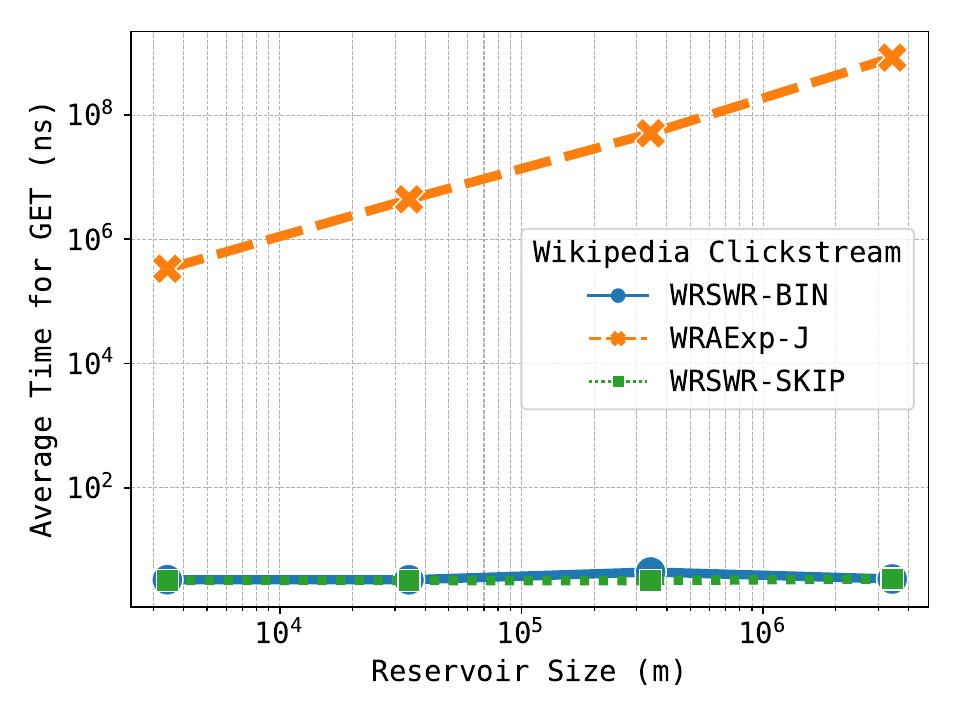}
        
        \vspace{-0.04cm}
        \caption{\small \texttt{Get}.}
        \label{fig:semisynthetic_experiment_get}
    \end{subfigure}
    
    \vspace{-0.3cm}
    
    \caption{\small Average time (ns) over 100 executions for \texttt{Add} (left) and \texttt{Get} (right) operations vs. reservoir size ($m$). Stream of 34M items from Wikipedia Clickstream dataset.}
    \label{fig:semisynthetic_experiment}
    
    \vspace{-0.4cm}
\end{figure}
\spara{Experiments on Synthetic Data.} Figure~\ref{fig:synthetic_experiment} illustrates the comparative results on a synthetic stream of length $N=10^7$. The analysis covers three weight distributions: decreasing ($w_i=i^{-1}$), constant ($w_i=1$), and increasing ($w_i=i$) for all $i \in [N]$. 
Figures~\ref{fig:synthetic_experiment_add_dec}, ~\ref{fig:synthetic_experiment_add_const}, and~\ref{fig:synthetic_experiment_add_inc} illustrate the \texttt{add} performance of \texttt{WRSWR-SKIP} (dotted line) across all three weight distributions. 
While \texttt{WRSWR-SKIP} has an execution time comparable to \texttt{WRAExp-J} (dashed line) for small $m$, its cost scales more slowly as $m$ increases.
The execution time of \texttt{WRAExp-J} increases significantly with $m$, eventually surpassing that of \texttt{WRSWR-BIN} (solid line) at the largest reservoir size ($m=10\% \, N$). Furthermore, \texttt{WRSWR-BIN} consistently exhibits higher execution times than \texttt{WRSWR-SKIP} across all tested reservoir sizes.
The \texttt{add} performance gap between \texttt{WRSWR-SKIP} and \texttt{WRAExp-J} can be attributed to update complexity. \texttt{WRSWR-SKIP} uses constant-time array updates, whereas \texttt{WRAExp-J} requires a priority queue. The resulting $O(\log{m})$ update cost contributes to widen the gap as $m$ increases.

Figures~\ref{fig:synthetic_experiment_get_dec}, ~\ref{fig:synthetic_experiment_get_const}, and~\ref{fig:synthetic_experiment_get_inc} show the results for the \texttt{get} operation, which highlight a clear distinction in algorithmic complexity. \texttt{WRSWR-SKIP} and \texttt{WRSWR-BIN} show nearly identical, constant-time performance, with execution times remaining flat regardless of reservoir size. This indicates an $O(1)$ retrieval, consistent with the theory. \texttt{WRAExp-J}, however, shows a clear linear increase in \texttt{get} time relative to $m$, which is consistent with its $O(m)$ retrieval complexity.

\spara{Experiments on the Wikipedia Clickstream data.} Figure~\ref{fig:semisynthetic_experiment} shows the algorithms' performances on the real-world Wikipedia Clickstream dataset, which contains 34 million items\footnote{December 2024, English Wikipedia pages: ~\url{https://dumps.wikimedia.org/other/clickstream/2024-12/clickstream-enwiki-2024-12.tsv.gz}}.
In this stream, the item element is the title of the requested Wikipedia article, and the weight is the number of clicks from another Wikipedia page or HTML page class\footnote{\url{https://meta.wikimedia.org/wiki/Research:Wikipedia_clickstream}}. 
As shown in Figure~\ref{fig:semisynthetic_experiment_add}, \texttt{WRSWR-SKIP} (dotted line) maintains the lowest execution time for the \texttt{add} operation. \texttt{WRSWR-BIN} (solid line) exhibits the highest cost for most of the range, but its execution time scales better than that of \texttt{WRAExp-J} (dashed line). \texttt{WRAExp-J} shows a steep increase in cost, becoming the least performant at the largest reservoir size. 
As seen in Figure~\ref{fig:semisynthetic_experiment_get}, the \texttt{get} performance on real-world data confirms the previous findings. \texttt{WRSWR-SKIP} and \texttt{WRSWR-BIN} deliver constant $O(1)$ retrieval, while the cost for \texttt{WRAExp-J} scales linearly with the reservoir size.

\section{Conclusion}
\label{sec:conclusion}
In this paper, we introduce \texttt{WRSWR-SKIP}, a novel one-pass algorithm for weighted sampling with replacement from data streams of unknown population size.
We formally prove the correctness of \texttt{WRSWR-SKIP} and analyze its efficiency, showing that the reservoir update operation generates $O\left(m \log \frac{W_N}{w_1}\right)$ random variates in expectation. Furthermore, \texttt{WRSWR-SKIP} performs sample extraction optimally in $O(1)$ time, as its reservoir is continuously maintained as an unbiased sample and requires no post-processing.
Experimental results from both synthetic and real-world datasets consistently demonstrate that \texttt{WRSWR-SKIP} outperforms the baselines.

\texttt{WRSWR-SKIP} provides an efficient and theoretically sound solution for weighted reservoir sampling with replacement, making it highly suitable for streaming applications where both fast processing and immediate sample retrieval are critical.



\bibliographystyle{ACM-Reference-Format}
\bibliography{biblio}
\end{document}